\documentclass[aps,twocolumn,showpacs,preprintnumbers,amsmath,amssymb]{revtex4}
\usepackage[dvips]{graphicx}
\usepackage{dcolumn}
\usepackage{amsmath}
\usepackage{amssymb}
\usepackage{units}
\relax

\usepackage{color}
\usepackage{ulem}

\begin{document}

\title{Hydrogen-induced ferromagnetism in ZnO single crystals investigated by Magnetotransport}

\author{M. Khalid}
\email{m.khalid@physik.uni-leipzig.de}
\author{P. Esquinazi}
\address{ Division of
Superconductivity and Magnetism, Institute for Experimental
Physics II, University of Leipzig, D-04103 Leipzig, Germany}

\date{\today}

\begin{abstract}
We investigated the electrical and magnetic properties of low-energy hydrogen-implanted ZnO single
crystals with hydrogen concentrations up to $\sim 3$ at.\% in the
first 20~nm surface layer between
10~K and 300~K. All samples showed clear ferromagnetic
hysteresis loops at 300~K with a saturation magnetization up to
$\simeq$ 4~emu/g. The measured anomalous Hall effect agrees  with
the hysteresis loops measured by superconducting quantum interferometer device
magnetometry. All the H-treated ZnO crystals
exhibited a negative magnetoresistance up to the room temperature. The
relative magnitude of the anisotropic magnetoresistance
reaches 0.4~\% at 250~K and 2~\% at 10~K, exhibiting an anomalous,
non-monotonous behavior and a change of sign below 100~K. All the experimental data
indicate that hydrogen atoms alone in a  few percent range
trigger a magnetic order in a ZnO crystalline state. Hydrogen implantation turns out to be a
simpler and effective method to generate a magnetic order in ZnO,
which provides interesting possibilities for future applications
due to the strong reduction of the electrical resistance.
\end{abstract}

\maketitle
\section{Introduction}

Defect-induced magnetism (DIM) appears to be a general phenomenon
observed in nominally non-magnetic solids starting from the
archetype graphite\cite{pabloprl03,ohldagnjp} to several oxides
like ZnO, pure or  doped with non-magnetic elements
\cite{Pan07,Bhos08,pot08,Xu08,kha09,kha10}, HfO$_2$
\cite{Venka04}, TiO$_2$ \cite{duha05,Hong06}, SrTiO$_3$
\cite{kha10,pot11}, SrO:N \cite{Elf07} as well as Si-based
samples\cite{liu11}, to mention only  a few examples from a large
number being reported nowadays (for recent reviews on this subject
see Refs.~\onlinecite{sto10,vol10,yaz10,and10}). Experimental
facts demonstrate that defects, like vacancies, without or with
the presence of non-magnetic ad-atoms, play a main role in
triggering magnetic order in these systems. Recently, room
temperature ferromagnetism was reported in Cu-doped ZnO films,
investigated by soft x-ray magnetic circular dichroism
\cite{Herng10}. The results of this study strengthens the
existence of the DIM phenomenon in general and in ZnO in particular.

ZnO is a wide band gap semiconductor, which crystallizes in the
hexagonal wurtzite, zincblende and rocksalt structures. However,
hexagonal wurtzite is the most intensively studied crystal
structure of ZnO because of its potential applications in the
field of spintronics, transparent electronics, piezoelectricity,
optoelectronics, etc. Hydrogen is one of the most abundant and
unavoidable impurities in ZnO. The presence of hydrogen can
influence the electrical and the magnetic properties of ZnO. The
role of hydrogen in enhancing the ferromagnetism in
3d-transition metal-doped ZnO was recently studied
experimentally \cite{Sing10} and theoretically\cite{Ass10}.
On the other side there are theoretical
studies reporting on the possibility of room temperature
ferromagnetism due to hydrogen adsorption at the surface of
ZnO\cite{San10,Liu09}. We note however that there is a lack of
 experimental and theoretical studies on the possibility of
hydrogen-induced magnetic order inside the crystalline structure
of pure ZnO.

In this work we are interested on DIM in ZnO crystals triggered
through the implantation of hydrogen at low energies and its influence
on the magnetotransport. We report on a detailed experimental
study that demonstrates how the intentional doping of hydrogen
in the percent range and at low-enough energies influences
substantially the electrical, magnetic and magnetotransport
properties of ZnO single crystals. In particular, we show in this
report a hydrogen-induced anisotropic magnetoresistance (AMR) as
well as the anomalous Hall effect (AHE) in H-ZnO single crystals.
Therefore, the existence of hydrogen-induced ferromagnetism in
H-ZnO samples is supported not only by the usual magnetization
data taken with a superconducting quantum interference device
(SQUID)\cite{khanjp} but also from magnetotransport measurements.

\section{Experimental details}

Hydrothermally grown ZnO (0001) single crystals of dimensions
(6$\times$6$\times$0.5) mm$^3$ with both sides polished were
supplied by CrysTec GmbH. The ZnO samples were exposed to remote
hydrogen dc plasma for different time intervals in a parallel-plate system. The voltage
difference between the two plates was kept at 1~kV. The samples
were mounted on a heater block held at a fixed temperature of
$400\,^{\circ}{\rm C}$ and they were placed $\sim$ 100~mm down
stream from the plasma with a bias voltage of $\sim$~-330~V. A
bias current of $\sim$ 50~$\mu$A was measured during the plasma
treatment. The pressure in the chamber during the process was
maintained around 1~mbar. Three samples H-1, H-2 and H-3 for time
intervals 30, 60 and 90~min, respectively, were treated in the
H-plasma chamber. The implantation depth (for the chosen energy)
as well as for the concentration characterization analysis (see
below)  were estimated using SRIM\cite{Zie85}. From this Monte
Carlo simulation program we estimate a penetration depth of 20~nm
for the implanted hydrogen atoms. As experimentally shown in
Ref.\cite{khanjp} the main ferromagnetic signal comes from
this near surface region, in agreement with the estimates.

Nuclear reaction analysis (NRA) was used to determine the hydrogen
concentration in ZnO crystals before and after hydrogen plasma
treatment \cite{Lan95}. The NRA has a depth resolution of $\sim$
5~nm with an average error in the concentration of 0.02\%. The
hydrogen concentration in ZnO crystals measured by NRA before and
after remote hydrogen treatment was found to be 0.14$\pm$0.03 and
0.64$\pm$0.07 at.$\%$ in the first 200~nm from the surface,
respectively\cite{Anw10}. From this concentration analysis we
conclude that the first 20~nm near surface region should have a
hydrogen concentration of the order of $\sim 3~$at.\% for sample
H-3. The hydrogen concentrations for samples H-1 and H-2 are
$\sim 1~$at.\% and $\sim 1.8~$at.\%, respectively.

We performed Particle Induced X-ray Emission (PIXE) measurements
to analyze the concentration of magnetic impurities in the H-ZnO
samples. There was no significant difference in the Fe
concentration ($< 60~$ppm) before and after H-plasma treatment.

The magnetization of the ZnO single crystals before and after the
plasma treatment  was determined with a SQUID. The
magnetotransport measurements were performed in an Oxford cryostat
with a magnetic field up to 8~T and a rotating sample holder
allowing us to measure the resistance at different angles between
magnetic field and the input current. The electrical contacts on
the samples were prepared using silver paste in a Van der Pauw
configuration. The $I/V$ characteristics  were measured to check
for deviations from the ohmic behavior. All the transport data
presented in this paper were taken in a linear, ohmic regime. The
resistance was measured with an ac resistance bridge with a
relative resolution of 0.01\%.

\section{results and discussion}
\subsection{Magnetization Measurements}

\begin{figure}
\begin{center}
\includegraphics[width=0.8\columnwidth]{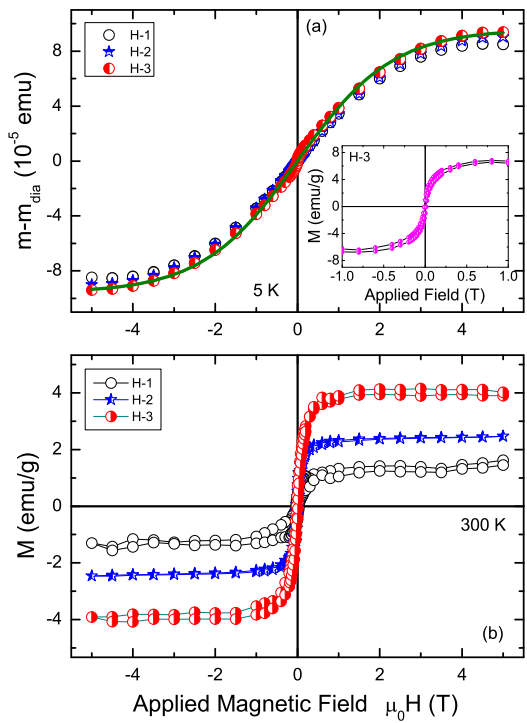}
\caption{(a) Magnetic  moment of the three H-ZnO crystals as a
function of magnetic field at 5~K. The inset shows a ferromagnetic
contribution which is left after subtracting a paramagnetic one
obtained by using the Langevin function.(b) Ferromagnetic magnetization
of the same crystals at 300~K assuming a homogeneously distributed
ferromagnetic mass at the first 20~nm near surface
region\protect\cite{khanjp}. A diamagnetic slope was subtracted
from the data. Note that the magnetization of H-ZnO crystals
increases with H-concentration.} \label{fig:1}
\end{center}
\end{figure}

Figure \ref{fig:1}(a) shows the magnetic moment after subtraction
of the linear diamagnetic contribution vs. applied field measured
at 5~K for all three samples with the SQUID. This signal is
composed by two contributions. The main one is paramagnetic and
follows the usual Brillouin or Langevin function. If one subtracts
it from the data a small ferromagnetic contribution still remains,
as can be seen in the inset of Figure \ref{fig:1}(a) for sample
H-3.

At 300~K the paramagnetic contribution is negligible and the
measured signal is given by the addition of the diamagnetic
plus the ferromagnetic one of the near surface region. After
subtraction of the diamagnetic contribution the magnetization
coming from the ferromagnetic part was calculated assuming a
ferromagnetic mass homogeneously located  at the first 20~nm near
surface region\cite{khanjp}, see Fig.~\ref{fig:1}(b). We see that
the magnetization increases with H-implantation as reported in
Ref.~\onlinecite{khanjp} and also decreasing temperature, compare
the results of sample H-3 in Fig.~\ref{fig:1}(b) with those in the
inset of Fig.~\ref{fig:1}(a). The ferromagnetic magnetization per
total volume of a virgin, untreated ZnO crystal of the same type
as used here, is of the order of $10^{-4}~$emu/g\cite{kha10}. For
comparison and taking into account that a similar near surface
region in nonmagnetic oxide crystals could be the source for the
ferromagnetic signals\cite{kha10}, the untreated ZnO crystal would
have a saturation ferromagnetic magnetization half of that of
sample H-1.

All samples exhibit a coercivity of 18-20~mT at room
temperature. The saturation magnetization increases by increasing
the H-concentration and reaches $\simeq$ 4~emu/g for sample H-3.
The ferromagnetic magnetization at 5~K increases by $\sim 50$\% of
its value at room temperature (see inset in Fig. 1(a)). From the
measurement of the ferromagnetic remanent magnetic moment we
estimate a Curie temperature of $450 \pm 25$~K~\cite{khanjp}.

Recently, N. Sanchez et al.\cite{San10} have theoretically shown
that atomic hydrogen adsorbed on the Zn-ZnO (0001) surface can
form strong H-Zn bonds and lead to a metallic surface with a net
magnetic moment of 0.5~$\mu$$_B$ per H-atom. The obtained
magnetization values indicate  that our H-3 sample, for example,
would have a net magnetic moment of 0.2~$\mu$$_B$ per hydrogen
atom. This estimate is obtained taking into account the amount of
H implanted. The difference between the two estimates may indicate
that the assumed ferromagnetic mass is larger than the true one.
This appears plausible because the hydrogen atoms are not
necessarily homogeneously distributed inside its penetration
depth. We believe that in the first 20~nm depth we have regions
where the magnetic order is less developed and therefore one tends
to overestimate the ferromagnetic mass. This picture of a rather
inhomogeneous mixture of magnetic and non-magnetic regions is of
importance to interpret the transport data, as we shall discuss
below.

\subsection{Resistivity Measurements}

\begin{figure}
\begin{center}
\includegraphics[width=0.9\columnwidth]{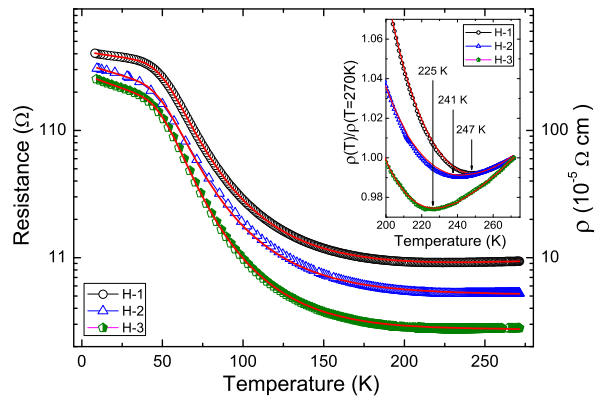}
\caption{Resistance of the three H-ZnO samples as a function
of temperature at a zero applied field. At the right $y-$axis we show the resistivity
estimated taken into account 20~nm implanted thickness of the
single crystal. The inset blows out the high temperature part
where the resistance shows a qualitative change in its temperature
dependence. The observed minimum shifts to the low
temperatures by increasing the hydrogen concentration. The red
solid lines show the fits of the data to Eq.~(2). The obtained values for $R_1$, $R_2$ and
$R_3$ are (0.022, 0.014, 0.015) $\Omega$, (0.587, 0.612, 0.365) $\Omega$ and
(197, 143, 169) $\Omega$ for samples (H-1, H-2, H-3), respectively. The rest
resistance for all samples is $R_0$ $\le$ 10$^-$$^6$ $\Omega$.}
\label{fig:2}
\end{center}
\end{figure}

We measured the temperature dependence of the resistance
of the three H-ZnO samples from 10~K to 270~K, see
Fig.~\ref{fig:2}. The resistance decreases increasing
H-concentration, therefore it is reasonable to assume that the
resistivity of the implanted part is much smaller than the
resistivity from the rest of the single crystal, as has been
already reported \cite{Thom54}. This assumption is supported by the
direct comparison of the estimated values of the resistivity --
assuming conduction within the 20~nm implanted thickness, right
$y-$axis in Fig.~\ref{fig:2} -- with the resistivity of the virgin
ZnO single crystal, which is already at room temperature several
orders of magnitude larger\cite{Bar10}.

Before discussing in detail the observed behavior  of the
resistivity, we would like to describe shortly what is known about
the hydrogen contribution to the formation and/or modification of
the electronic band structure in ZnO. Hydrogen forms shallow donor
states in bulk ZnO and is regarded as a source for n-type
conductivity. These shallow donor states are formed approximately
30-60~meV below the conduction band. Upon doping level the donor
energy states can be dispersed into an impurity band because of
the Coulomb fields arising from the compensating acceptors and
ionized donors\cite{Hung54}. This impurity band could further
split into two bands, i.e. a lower band I and an upper band I$^-$,
which are formed with single charged donors and
neutral donors, respectively\cite{Nish65,Shk84}.\\

As we have already noted in section 3A, we assume that the
implanted H$^+$-ions are not homogeneously distributed in H-ZnO
samples. Then, we may have hydrogen-rich metallic regions with resistance $R_m(T)$ = $R_0$ + $R_1 T$,
embedded in a doped semiconducting matrix
with resistance $R_s(T) = R_2 \exp(\Delta E/ 2k_B T)$ ($k_B$ is
the Boltzmann constant). $R_0$, $R_1, R_2$ are free parameters as
well as the activation energy $\Delta E$ which will be obtained by fitting the
experimental data. Note that we consider a linear $T$ dependence
for the metallic part in all the temperature range. The metallic
region contributes mainly at high enough temperatures, see inset
in Fig.~\ref{fig:2}, and therefore it is unnecessary to assume a
more complicated $T-$dependence that may be applicable at
temperatures $T < 100~$K. We consider that these two
contributions, the metallic- and semiconducting-like are in
series. Due to the implantation distribution curve \cite{khanjp}, it is clear
that below $\sim 20~$nm a third intermediate region should exist that
contributes with a resistance $R_h(T)$ in parallel to the other
two.The best fits have been achieved by assuming a variable range hopping (VRH)-like mechanism as:
\begin{equation}
R_h =  R_3 \exp\left(\frac{E_{nn}}{T}\right)^{1/5}\,.
\end{equation}
Where $R_3$ is a free parameter and $E_{nn}$ is a hopping energy. The
total resistance will be given then by

\begin{equation}
R(T) = \left[\left(R_h(T)\right)^{-1}+\left(R_m(T)+R_s(T)
\right)^{-1}\right]^{-1}\,. \label{1}
\end{equation}

One can also take into account a fourth parallel resistance
contribution arising from the pure ZnO single crystal below a
$\sim 100~$nm thick layer. However, the resistance of such pure
ZnO crystal at all  temperatures is in the range of M$\Omega$ and
therefore its contribution to the total resistance is negligible.
The fittings to the data of the three crystals are reasonably good as we
can see in Fig.~\ref{fig:2}. The activation energy obtained from
the fittings for the three H-ZnO single crystals is 60$\pm$2~meV.

Qualitatively the observed temperature dependence of the
resistivity is rather simple to understand. At temperatures below
50~K, the resistance of the semiconducting contribution $R_s$ is
larger than the resistance  $R_h(T)$, becoming this last the
dominant transport mechanism\cite{Say10} because the thermal
energy is not enough to excite the electrons from the upper
impurity band I$^-$ to the conduction band. The hopping energy
obtained from the fitting of the experimental data is $E_{nn}
\simeq 3\pm 0.5$~meV. As the temperature increases the resistance
of the H-ZnO samples decreases following a semiconducting behavior
with en effective activation energy $\Delta E \simeq 60$~meV. The
larger the hydrogen doping the lower is $R_m(T)$ and therefore the
lower is the temperature of the minimum, see inset in
Fig.~\ref{fig:2}. Although with this simple model we can
understand qualitatively the behavior measured in the resistance
as a function of temperature and hydrogen concentration, it does
not provide us with a clear hint about the regions that contribute
to the magnetic signal. If the magnetic order is confined mostly
within the first 20~nm surface region\cite{khanjp} we expect then
that either the metallic or the semiconducting regions or even an intermediate
region between these two and the VRH part, contributes
to the ferromagnetic signal. As the semiconducting
contribution overwhelms in a wide temperature range the metallic
one, it appears plausible that this one may be responsible for the
magnetoresistance behavior we have observed in H-ZnO single crystals.

\subsection{Charge Carriers}
\begin{figure}
\begin{center}
\includegraphics[width=0.8\columnwidth]{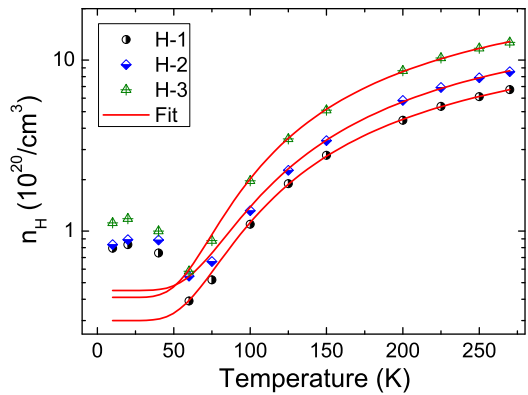}
\caption{\label{fig:3} (a) Carrier density of the three H-ZnO
single crystals as a function of temperature. The carrier density
shows an anomalous behavior around $\sim 50$~K. The solid lines are
fit to the expression given in Eq.(~\protect\ref{1a}).}
\end{center}
\end{figure}

We performed Hall measurements  in a Van der Pauw configuration in
order to obtain the charge carriers density of our H-ZnO samples.
Our Hall measurements  confirm that the conductivity in H-ZnO
single crystals is n-type.  We note that the estimated carrier
concentration is an effective one obtained using the simplest
expression $n=1/R_He$, where $R_H$ is the Hall resistance,
assuming that only the sample volume of the first 20~nm H-rich
surface layer of the ZnO single crystals contributes. In case
electrons and holes would contribute with different densities and
scattering rates then the two-band model is necessary to obtain
the carrier densities. The use of its equations, however, implies
the introduction  of  free, unknown parameters, like the
scattering rates. In order to facilitate the comparison of our
data with literature values  we prefer to discuss the carrier
density from the Hall data  stressing that the measurable quantity
is $R_H$. The temperature dependent of the carrier density $n_H$
is shown in Fig.~\ref{fig:3}.

The carrier concentration of samples H-1, H-2 and H-3  at room
temperature are 6.72 $\times$10$^2$$^0$cm$^-$$^3$, 8.56
$\times$10$^2$$^0$cm$^-$$^3$ and 1.27
$\times$10$^2$$^1$cm$^-$$^3$. These values are comparable to those
found in, e.g., Ga-doped ZnO system\cite{Yama07}. Above $T \sim 50~$K the carrier density increases with temperature
following the equation:
\begin{equation}
n_H = a + b\exp\left(\frac{\Delta E}{2 k_BT}\right)\,, \label{1a}
\end{equation}
where $a,b$ are free parameters. The activation energy obtained
from the fits is $\Delta E = 60 \pm$2~meV, in agreement
to the activation energy values obtained from the resistivity
measurements, see Fig.~\ref{fig:2}. As expected, the carrier
concentration increases with hydrogen concentration. The increase
in $n_H$ between the samples agrees roughly
with the estimated increase in the hydrogen concentration.

At temperatures $T \lesssim 50~$K, $n_H$ increases with a decrease in
temperature, see Fig.~\ref{fig:3}. This is an anomalous behavior
that appears to be related to the change of the main contribution
to the measured resistance, i.e. from the semiconducting region
above 50~K to the VRH one below it, see Fig.~\ref{fig:2}. In
this case it might be that the simple relation to estimate $n_H(T
< 50$K) from the Hall resistance is not adequate and a more
complicated equation for the Hall signal of a material with two
contributions in parallel should be used\cite{ash}. We note that
the anomaly at $T \sim 50~$K  is observed in all the magnetotransport properties we have measured,
as we will show in the following sections.

\subsection{Magnetoresistance Measurements}\label{mr}

\begin{figure}
\begin{center}
\includegraphics[width=0.8\columnwidth]{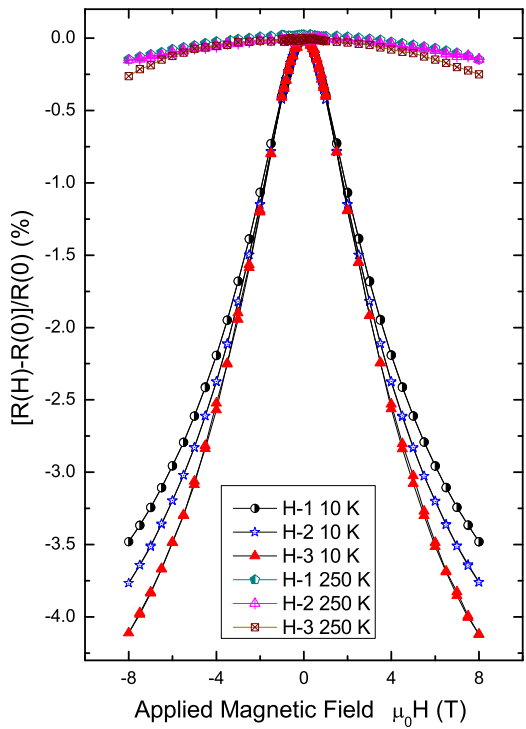}
\caption{\label{fig:4}  Magnetoresistance (in percentage) as a
function of applied magnetic field at 10~K and 250~K for the three
H-ZnO samples. A clear correlation between magnetoresistance and
H-concentration is observed. The magnetoresistance increases with
H-concentration.}
\end{center}
\end{figure}

Apart from the magnetization measurements a further and important
way to check for the existence of magnetic order is through the
measurement of the magnetotransport properties. Unlike
magnetization measurements, magnetotransport properties are much
less sensitive to magnetic impurities, in case they remain below
$\sim 0.1\%$, and in general they reflect intrinsic
characteristics of the sample. In this section we discuss the
longitudinal magnetoresistance of H-ZnO where the magnetic field
is applied parallel to the input current as well as to the sample
main plane. The longitudinal magnetoresistance for the three H-ZnO
samples measured up to 8~T at 10~K and 250~K is shown in
Fig.~\ref{fig:4}. The magnetoresistance is defined as
$[[R(H)-R(0)]/R(0)]$ where $R(H)$ and $R(0)$ are the resistances
with and without an applied magnetic field, respectively.

All samples show a negative magnetoresistance at  all temperatures
and magnetic fields applied parallel to the main plane of the
samples. The negative magnetoresistance has been observed in
several other ZnO systems that show some kind of magnetic
order\cite{Kum10,Shi04}.

\begin{figure}
\begin{center}
\includegraphics[width=0.8\columnwidth]{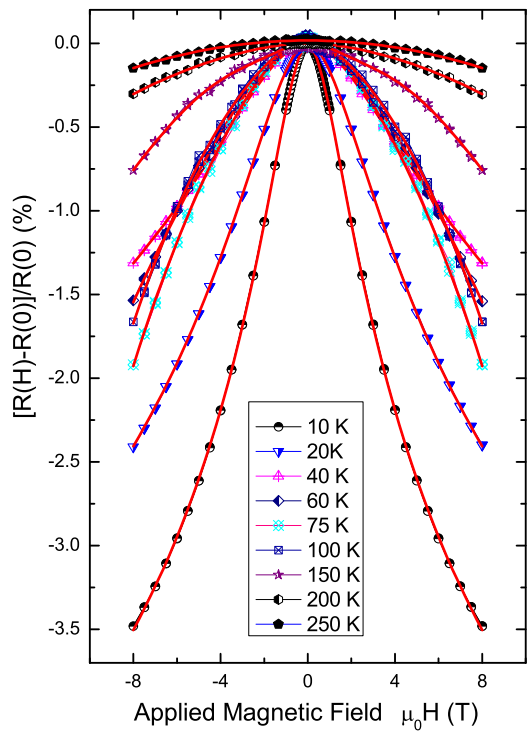}
\caption{\label{fig:5}  Magnetoresistance of sample H-1 (the
 other two samples show a similar
dependence) as a function of applied magnetic field measured at
several temperatures. The magnetoresistance shows a negative
temperature dependence but with  an anomaly around 50~K. The solid
(red) lines through the data points are fits to Eq.~(\ref{2}).}
\end{center}
\end{figure}

Figure~\ref{fig:5} shows the temperature dependence of the
magnetoresistance for the H-1 sample. Similar temperature
dependent magnetoresistance is observed for the other two samples.
It is clear from Fig.~\ref{fig:5} that the magnetoresistace of
H-ZnO decreases in general with temperature. However, its field
curvature at $T \sim 50$~K changes and from 50~K to 75~K the magnetoresistance
increases with temperature. The decrease in the magnetoresistance
with temperature is expected because the magnetization at
saturation of these samples also  decreases with temperature, see
Fig.~\ref{fig:1}. As we have observed in the carrier
concentration, the anomalous behavior in the magnetoresistance
around 50~K might be related to the change of the main
contribution to the resistance. Taking into account that the main
contribution to the magnetic signal comes from the 20~nm surface
contribution, the variable range hopping part and the magnetic
semiconducting part might have a common interface, which shows
magnetic order and contributes to the magnetoresistance at low
temperatures.

To elucidate our experimental results for the magnetoresistance we
use a model proposed by Khosla and Fischer\cite{Kho70} that
combines negative and positive magnetoresistances in
semiconductors  taking into account a third-order expansion of the
$s-d$ exchange Hamiltonian. The semiempirical formula is:

\begin{equation}
\frac{\Delta\rho}{\rho_0}=
-a^2\ln(1+b^2B^2)+\frac{c^2B^2}{1+d^2B^2}\,,
\label{2}
\end{equation}
where $c$ and $d$ depend on the conductivity and the carrier mobility, respectively.
We will consider these as free parameters, and
\begin{equation}
a^2 = A_1J\rho_F[S(S+1)+\langle{M^2}\rangle],
\label{2a}
\end{equation}

\begin{equation}
b^2 =
\left[1+4S^2\pi^2\left(\frac{2J\rho_F}{g}\right)^4\right]\frac{g^2\mu^2}{(\alpha
kT)^2}\,,
\label{2b}
\end{equation}

where $\mu$ is the mobility, $g$ is the Land$\acute{e}$ $g$-factor, $\alpha$ is a numerical constant. The fitting
parameters $a$ and $b$ in Eq.~(\ref{2}) depend on several factors
such as a spin scattering amplitude $A_1$, the exchange integral
$J$, the density of states at the Fermi energy $\rho_F$, the spin
of the localized magnetic moments $S$ and the average
magnetization square $\langle{M^2}\rangle$. The negative first term in Eq.~(\ref{2}) is
attributed to a spin dependent scattering in third order $s-d$
exchange Hamiltonian while the positive part (second term in the expression of Eq.~(\ref{2}))
takes into account field induced changes due to the
two, $s$ and $d$, conduction bands with different conductivities.

\begin{figure}
\begin{center}
\includegraphics[width=0.9\columnwidth]{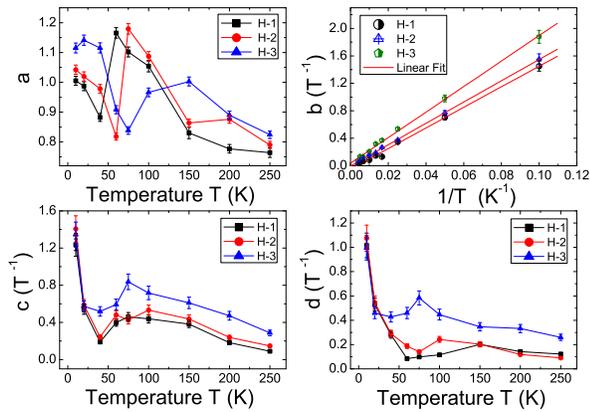}
\caption{\label{fig:6} Temperature dependence of the fitting parameters $a$, $b$, $c$ and $d$ of three H-ZnO samples.}
\end{center}
\end{figure}

The fits of the experimental data to Eq.~(\ref{2}) are shown in
Fig.~\ref{fig:5}. The data can be well fitted with
this model at all measured temperatures. All four fitting
parameters show different temperature dependence (see Fig.6). Note that the
positive magnetoresistance in our samples is compensated by the
large negative magnetoresistance contribution. Therefore, the
uncertainty of the parameters $c$ and $d$ is rather large because
they are not independent of the fitting procedure. Therefore, we
 concentrate on the negative scattering contributions $a$ and $b$.
 The fitting parameter $b$ for the three H-ZnO samples are plotted
as a function of inverse temperature in Fig.~\ref{fig:6}. The
parameter $b$ shows a linear dependence in good agreement with
theory. We found that the parameter $a$ is almost temperature
independent and increases with H-concentration
in the temperature range 50 K $\ge$ T $\ge$ 100 K indicating that the sample with higher
H-concentration (H-3) is more magnetic, in agreement with the
SQUID data shown in Fig.~\ref{fig:1}.

Parameters $a$ and $b$ defined in Eq.~(\ref{2a}) and Eq.~(\ref{2b}) respectively, are used in order to obtain
the values of $J\rho_F$ and $A_1$. The values of $J\rho_F$ and $A_1$ obtained from
the experimental data are 0.56 and 0.14 for $S$ = 1/2 and $\mu$ = 36 cm$^2$/V.s at 10 K, respectively.
The values of $J\rho_F$ and $A_1$
for $S$ = 3/2 are 0.33 and 0.12, respectively. The value of
$J\rho_F$ = 0.33 for $S$ = 3/2 in H-ZnO is similar to the one obtained in CdS system
$J\rho_F$ = 0.4 \cite{Kho70}. These results strongly suggest the contribution of $s$-$d$ interaction in the H-ZnO system.

\subsection{Anisotropic Magnetoresistance}\label{amr}

There are two other magnetotransport effects, which are observed
in our H-ZnO crystals and are worth mentioning. One of them is the
anisotropic magnetoresistance (AMR) effect. This effect represents
the change in the resistance of a ferromagnetic material with the
angle between the input current and applied field in plane. It is
commonly associated with the presence of a spin splitting of the
electronic band at the Fermi level and a finite spin-orbit ($L-S$)
coupling. The AMR arises in second order in the $L-S$ coupling, in
contrast to the magnetocrystalline anisotropy. In ferromagnetic
materials with $s-$ and $d-$bands the AMR is understood arguing
that the spin-orbit scattering increases the resistance by
allowing a spin-flip and through this the occupation of free
$d-$states in the corresponding spin dependent band. In general it
is expected that the resistance is larger when the applied field
is parallel to the current. We define the AMR amplitude as
\begin{equation}
\Delta R/R_{\rm avg} = \frac{|R(H_{\|})-R(H_{\bot})|}{
(R(H_{\|})+R(H_{\bot}))/2}.\label{eqmar}
\end{equation}
For polycrystalline ferromagnetic samples the change of the
resistance due to the AMR has the following angle dependence:
\begin{equation}
\frac{\Delta R}{R(H=0)}= A\cos^2\theta\,, \label{3}
\end{equation}
where $\theta$ is the angle between the current $I$ and the applied
field or magnetization direction (in saturation) and $A$ is a
constant that depends on the density of $d$-states at the Fermi
level, on the magnetization as well as on the sample quality.

\begin{figure}
\begin{center}
\includegraphics[width=0.8\columnwidth]{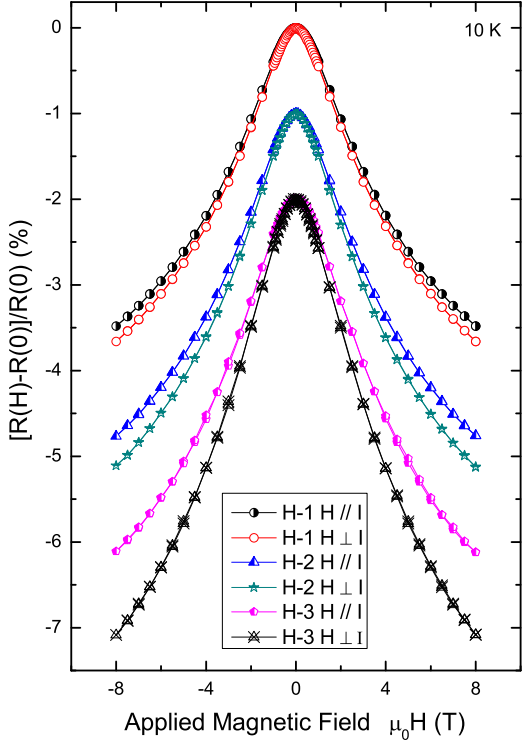}
\caption{\label{fig:7} Magnetoresistance of the H-ZnO samples at
10~K for two configurations of angles between current and the
applied magnetic field. For clarity the data curves of the samples H-2
and H-3 are shifted by a constant value.}
\end{center}
\end{figure}

Figure~\ref{fig:7} shows the field dependence of the
magnetoresistance  for $\theta$ = $0\,^{\circ}$ and $90\,^{\circ}$
at 10~K  for the  three H-ZnO samples. The angle $\theta$ is the
angle between the applied current and the magnetic field. We
observe a clear AMR effect in the three H-ZnO samples. We note
that the AMR effect was already measured in ZnO
 but doped with Co\cite{Lee06}. Our results show clearly that a fundamental
 property of ferromagnetic materials as the AMR can also be obtained by DIM in an oxide.
 The AMR effect in H-ZnO single crystals increases with
 H-concentration, see Fig.~\ref{fig:7}.

\begin{figure}
\begin{center}
\includegraphics[width=0.9\columnwidth]{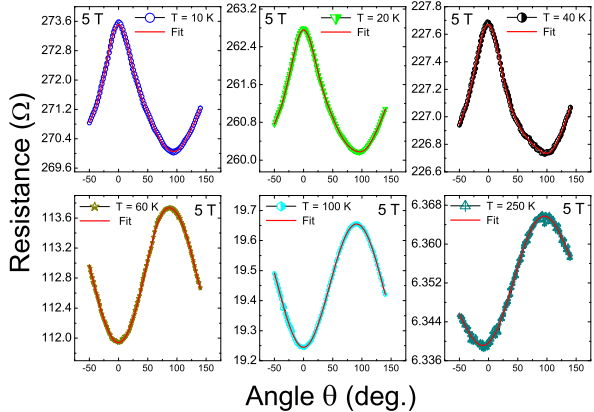}
\caption{\label{fig:8} Angular magnetoresistance of sample H-1 at
a constant field of 5~T at several temperatures. The red solid
lines are the fits to experimental data following
Eq.~(\protect\ref{4}).}
\end{center}
\end{figure}

However, for ferromagnetic single crystalline systems some
influence from the lattice anisotropy can be expected in the AMR
and in this case the angular-dependent magnetoresistance is
described by a Fourier series of $\cos(n\theta$) and
$\sin(n\theta$) as
\begin{equation}
\frac{\Delta R}{R(H=0)}= \sum_{n=1}^{n=\infty}s_n
\sin(n\theta)+\sum_{n=1}^{n=\infty}c_n \cos(n\theta)\,, \label{4}
\end{equation}
where the coefficients $s_n$ and $c_n$ are related to the Hall and
magnetoresistance contributions in the system.

\begin{figure}
\begin{center}
\includegraphics[width=0.9\columnwidth]{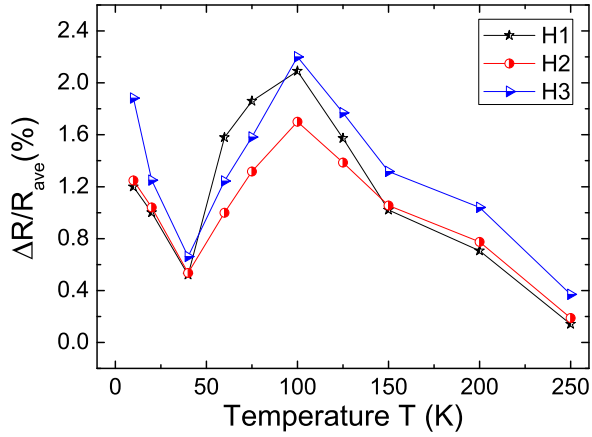}
\caption{\label{fig:9} Temperature dependence of the absolute
value of the AMR defined in Eq.~(\ref{eqmar}), for the
three H-ZnO samples. Note that actually the AMR changes sign at $T
\simeq 50~$K.}
\end{center}
\end{figure}

\begin{figure}
\begin{center}
\includegraphics[width=0.9\columnwidth]{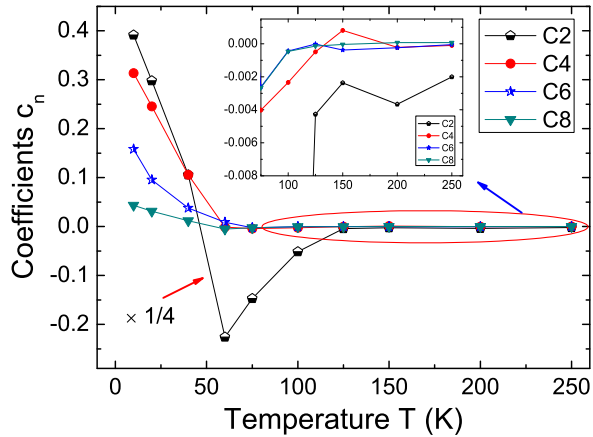}
\caption{\label{fig:10} Fitting coefficient $c_n$ as a function of
temperature obtained from the fits to Eq.~(\protect\ref{4}) to the
experimental data shown in Fig.~\protect\ref{fig:8}. The
values of the coefficient $c_2$ are divided by 4 to include them
for clarity with the other coefficients. Note that it changes sign at $\sim$ 50 K.}
\end{center}
\end{figure}

In order to study the AMR in more detail, we measured the angular
magnetoresistance at a constant field of 5~T at several
temperatures, see Fig.~\ref{fig:8}. The variation of the angle
between the magnetic field and the current ranges from $-50^\circ$
to $140^\circ$. This range of angle is necessary because we
expect a $180\,^{\circ}$ periodicity. There are several distinct
features in the experimental data shown in Fig.~\ref{fig:8}. (a)
At $T < 50~$K the AMR is higher at $\theta$ = $0\,^{\circ}$ (field
and current parallel to each other) than at $\theta$ =
$90\,^{\circ}$, in agreement with the usual behavior. (b) At
temperatures $T > 50~$K the AMR changes sign (or it shows a
$90^\circ$ shift in the angle dependence), see Fig.~\ref{fig:8}.
(c) Figure~\ref{fig:9} shows the AMR amplitude, see
Eq.(\ref{eqmar}), as a function of temperature for the three
samples. There is a clear anomalous increase with temperature
between 50~K and 100~K, in spite of the fact the the magnetization
at saturation decreases monotonously in all the temperature range
according to the SQUID measurements (not shown). Note that the magnitude of
the AMR in our H-ZnO samples at 250~K is $\sim$ 0.4 $\%$, a value
comparable to the AMR observed in, e.g., Co films\cite{barapl11}
or Co:Cu multilayered nanowires\cite{Tan08}. (d) The measured
angle dependence does not follow  Eq.~(\ref{3}) applicable for
polycrystalline materials and indicates that the magnetic contribution in our
H-ZnO single crystals comes from a single crystalline phase after
H-plasma treatment. The AMR curves obtained for all three H-ZnO
samples are fitted by Eq.~(\ref{4}) and the results of these fits
are shown in Fig.~\ref{fig:8} as (red) solid lines. The
experimental data can be fitted quite well at all temperatures
after expanding Eq.~(\ref{4}) up to eighth order. The coefficients
$s_n$ and $c_n$ are related to the antisymmetric (Hall) and
symmetric (magnetoresistance) contributions of the sample. The
values of the coefficients $s_n$ obtained from the fits are
negligibly small and therefore only the coefficients $c_n$ are shown
in Fig.~\ref{fig:10}. The major contributions to the AMR come from
the terms with $n = 2, 4, 6$ and 8 indicating that the action of
the Lorentz force on the mobile charges is not the source for the
observed AMR in H-ZnO samples.

As shown for the system ZnO-Cu with oxygen vacancies \cite{her10},
however, the influence of hydrogen cannot rule out the Zn-$d$
contribution to the magnetic order as well as
from the oxygen $p$-states. X-ray magnetic circular dichroism
measurements are necessary to obtain the required information on
the elements (and bands) contributions to the observed magnetic
order. In particular the origin of the clear anomalous behavior at
$50~K \lesssim T \lesssim 100$~K with the unexpected change of sign
of the AMR effect (factor C2, see Fig.10) requires further studies that go beyond a
magnetotransport characterization. We note, however, that a change of sign of the AMR (and
the thermopower $S$) at $T \sim 50~K$ has been observed for particular field directions
in U$_3$As$_4$ and U$_3$P$_4$ ferromagnetic single crystals \cite {Pio07}. The author interpreted
the abrupt change and sign inversion of AMR (and $S$) in the frame of spin-orbit coupling (SOC) and
a large sensitivity in energy of the spin density of states at the Fermi-level due to a spontaneous trigonal
distortion in magnetically ordered state.

\subsection{Anomalous Hall Effect} \label{ahe}

\begin{figure}
\begin{center}
\includegraphics[width=0.8\columnwidth]{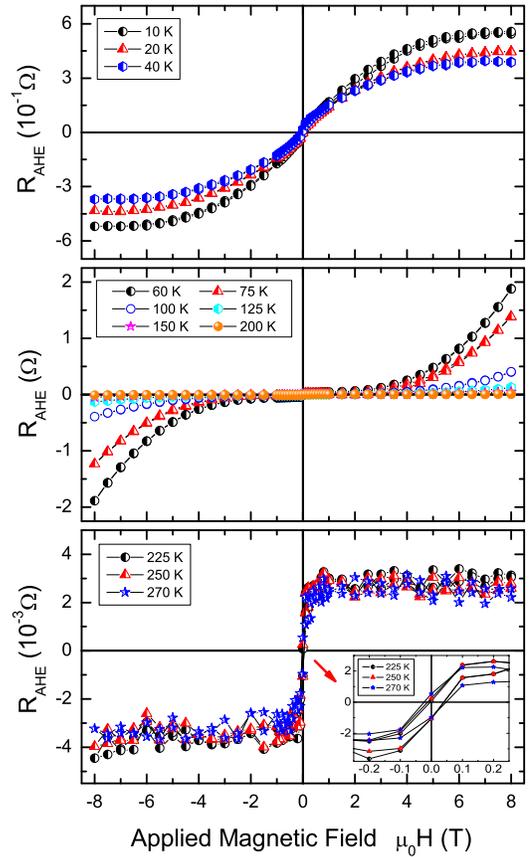}
\caption{\label{fig:11} Anomalous Hall resistance of the H-2
sample as a function of a magnetic field at several temperatures.
The linear background from the conventional Hall effect was
subtracted from the measured curves. Note that AHE shows an
anomalous change of field curvature above 50~K and it is nearly
temperature independent above 225~K.}
\end{center}
\end{figure}

The anomalous Hall effect (AHE) was reported in magnetic-ion doped
ZnO systems in the past\cite{Hsu08,Yan08,Qin07} but not yet in an
(magnetic-ion) un-doped ZnO systems and at room temperature. The
Hall resistance in ferromagnetic materials consists of two
contributions which are the ordinary Hall resistance (due to
Lorentz force) and the anomalous Hall resistance (due to an
asymmetric scattering in the presence of magnetic order) and can
be expressed by the following equation
\begin{equation}
R_{\rm Hall}= R_H(H)+R_{AHE}(M)\,, \label{5}
\end{equation}
where $R_{H}$ and the $R_{AHE}$ are the ordinary and anomalous Hall
resistances and $M$ is the magnetization. The dominant feature of
the Hall data in our H-ZnO samples is a linear dependence of
$R_{\rm Hall}$ with magnetic field with a negative slope due to
the ordinary contribution $R_H$. After subtracting $R_H(H)$ from
the measured data, an anomalous Hall effect contribution is
obtained for all three H-ZnO samples, as shown in
Fig.~\ref{fig:11} for sample H-2. Clear $s-$like loops with very
weak hysteresis are observed in $R_{AHE}(H)$ up to 300~K. The shape of
the loops at temperatures $ T < 50~$K indicates that the hydrogen
related paramagnetic centers dominate, in agreement  with SQUID
results, see Fig.~\ref{fig:1}(a). At intermediate temperatures $50
\lesssim T \lesssim 150$~K the behavior of $R_{AHE}(H)$ is anomalous
in the sense that it has a different field curvature without
saturation at large fields. Note that in the same temperature
range the carrier concentration, magnetoresistance and anisotropic
magnetoresistance behave also anomalously. At temperatures above
150~K the $R_{AHE}(H)$ curves follow the expected behavior for a
ferromagnet, see Fig.~\ref{fig:11}.
\begin{figure}
\begin{center}
\includegraphics[width=0.9\columnwidth]{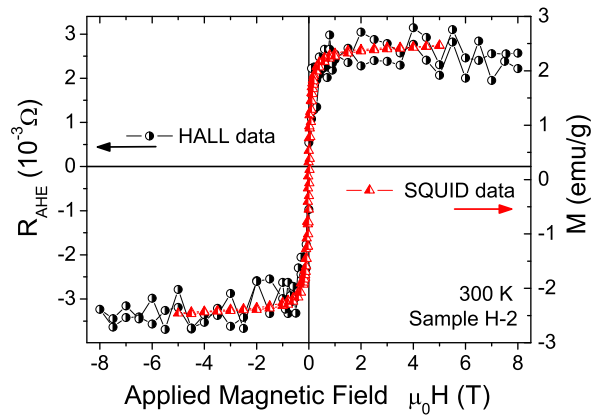}
\caption{\label{fig:12} A comparison of the SQUID and the
anomalous Hall resistance data at 300~K.}
\end{center}
\end{figure}

In order to further investigate whether the anomalous Hall effect
is mainly affected by the magnetization response of the samples we
compared the SQUID and $R_{AHE}(H)$ results. Figure ~\ref{fig:12}
shows the $M(H)$ and $R_{AHE}(H)$ loops at 300~K. Both curves are
very similar, which indicate that the AHE originates from
the intrinsic ferromagnetism of the H-ZnO samples.

\section{Conclusion}

We have studied the magnetic and magnetotransport properties of
H-implanted ZnO single crystals with different hydrogen
concentrations in the atomic percent range. Clear
ferromagnetic-like loops were observed in all three H-ZnO samples
at room temperature with a magnetization at saturation up to 4
emu/g. The Hall measurements confirmed the n-type transport
mechanism in H-ZnO. We observed a negative magnetoresistance in
all crystals and in the available temperature and magnetic field
range. The magnitude of the magnetoresistance increases with
H-concentration. We observed the anomalous Hall effect and
anisotropic magnetoresistance in the H-ZnO single crystals. The
magnitude of anisotropic magnetoresistance was found to be 0.4 \%
at 250 K, a value comparable to polycrystalline cobalt. The
anomalous Hall effect data showed a quantitative agreement to the
SQUID data for all three
H-ZnO single crystals, a fact that excludes impurities as the
origin of the observed ferromagnetism. The observation of
anisotropic magnetoresistance up to room temperature strongly
suggests the presence of a spin-splitted band with a non-zero
spin-orbit coupling in H-ZnO single crystals. At temperatures below 100 K,
anomalous behaviors in the magnetoresistance, anisotropic magnetoresistance, and
carrier density were observed that are apparently related to the change of the main
contribution to the measured resistance, i.e from semiconducting to VRH-like transport. We believe that our
findings would be useful for further understanding of defect
induced magnetism in ZnO as well as other oxide systems and could
be the starting point towards an efficient and reproducible way of
inducing ferromagnetism in ZnO systems.

\acknowledgments We gratefully acknowledge Prof. J. Weber and Dr. E. Lavrov
from the Technical University of Dresden for the support in
preparing the ZnO crystals in their laboratory. We thank Dr. M.
Ziese for fruitful discussions. This work was supported by the DFG
within the Collaborative Research Center (SFB 762) ``Functionality
of Oxide Interfaces''.


\end{document}